\documentstyle[psfig,general_cite]{mn}
\newif\ifAMStwofonts

\def\hi{H{\sc i} \,}
\def\lxlb{L$_X$/L$_B$ \,}
\def\eg{{\it e.g.\ }}
\def\ie{{\it i.e.\ }}

\ifoldfss
  \ifCUPmtlplainloaded \else
    \NewTextAlphabet{textbfit} {cmbxti10} {}
    \NewTextAlphabet{textbfss} {cmssbx10} {}
    \NewMathAlphabet{mathbfit} {cmbxti10} {} 
    \NewMathAlphabet{mathbfss} {cmssbx10} {} 
  \fi
  \ifAMStwofonts
    \ifCUPmtlplainloaded \else
      \NewSymbolFont{upmath} {eurm10}
      \NewSymbolFont{AMSa} {msam10}
      \NewMathSymbol{\upi}     {0}{upmath}{19}
      \NewMathSymbol{\umu}     {0}{upmath}{16}
      \NewMathSymbol{\upartial}{0}{upmath}{40}
      \NewMathSymbol{\leqslant}{3}{AMSa}{36}
      \NewMathSymbol{\geqslant}{3}{AMSa}{3E}

      \let\leq=\leqslant 
      \let\geq=\geqslant 
    \fi
  \fi
\fi 

\ifnfssone
  \newmathalphabet{\mathit}
  \addtoversion{normal}{\mathit}{cmr}{m}{it}
  \addtoversion{bold}{\mathit}{cmr}{bx}{it}
  \newmathalphabet{\mathbfit} 
  \addtoversion{normal}{\mathbfit}{cmr}{bx}{it}
  \addtoversion{bold}{\mathbfit}{cmr}{bx}{it}
  \newmathalphabet{\mathbfss} 
  \addtoversion{normal}{\mathbfss}{cmss}{bx}{n}
  \addtoversion{bold}{\mathbfss}{cmss}{bx}{n}
  \ifAMStwofonts
    \ifCUPmtlplainloaded \else
      \UseAMStwoboldmath
      \makeatletter
      \new@mathgroup\upmath@group
      \define@mathgroup\mv@normal\upmath@group{eur}{m}{n}
      \define@mathgroup\mv@bold\upmath@group{eur}{b}{n}
      \edef\UPM{\hexnumber\upmath@group}
      \new@mathgroup\amsa@group
      \define@mathgroup\mv@normal\amsa@group{msa}{m}{n}
      \define@mathgroup\mv@bold\amsa@group{msa}{m}{n}
      \edef\AMSa{\hexnumber\amsa@group}
      \makeatother
      \mathchardef\upi="0\UPM19
      \mathchardef\umu="0\UPM16
      \mathchardef\upartial="0\UPM40
      \mathchardef\leqslant="3\AMSa36
      \mathchardef\geqslant="3\AMSa3E

      \let\leq=\leqslant 
      \let\geq=\geqslant 
    \fi
  \fi
\fi 

\ifnfsstwo
  \DeclareMathAlphabet{\mathbfit}{OT1}{cmr}{bx}{it}
  \SetMathAlphabet\mathbfit{bold}{OT1}{cmr}{bx}{it}
  \DeclareMathAlphabet{\mathbfss}{OT1}{cmss}{bx}{n}
  \SetMathAlphabet\mathbfss{bold}{OT1}{cmss}{bx}{n}
  \ifAMStwofonts
    \ifCUPmtlplainloaded \else
      \DeclareSymbolFont{UPM}{U}{eur}{m}{n}
      \SetSymbolFont{UPM}{bold}{U}{eur}{b}{n}
      \DeclareSymbolFont{AMSa}{U}{msa}{m}{n}
      \DeclareMathSymbol{\upi}{0}{UPM}{"19}
      \DeclareMathSymbol{\umu}{0}{UPM}{"16}
      \DeclareMathSymbol{\upartial}{0}{UPM}{"40}
      \DeclareMathSymbol{\leqslant}{3}{AMSa}{"36}
      \DeclareMathSymbol{\geqslant}{3}{AMSa}{"3E}

      \let\leq=\leqslant 
      \let\geq=\geqslant 
    \fi
  \fi
\fi 

\ifCUPmtlplainloaded \else
  \ifAMStwofonts \else 
    \def\upi{\pi}
    \def\umu{\mu}
    \def\upartial{\partial}
  \fi
\fi
\def\etal{{\it et al. }}

\begin{document}

\title[
X--rays from Post--Mergers
]
{
The X--ray Emission in Post--Merger Ellipticals
}
\author[
Ewan O'Sullivan et al. 
]
{
Ewan O'Sullivan$^{1}$, Duncan A. Forbes$^{1,2}$  
and Trevor J. Ponman$^{1}$\\
$^1$School of Physics and Astronomy, 
University of Birmingham, Edgbaston, Birmingham B15 2TT \\
(E-mail: ejos@star.sr.bham.ac.uk)\\
$^2$Astrophysics \& Supercomputing, Swinburne University,
Hawthorn VIC 3122, Australia\\
\\
}

\pagerange{\pageref{firstpage}--\pageref{lastpage}}
\def\LaTeX{L\kern-.36em\raise.3ex\hbox{a}\kern-.15em
    T\kern-.1667em\lower.7ex\hbox{E}\kern-.125emX}

\newtheorem{theorem}{Theorem}[section]

\label{firstpage}

\maketitle

\begin{abstract}
The evolution in X--ray properties of early--type galaxies is
largely unconstrained. In particular, little is known about how,
and if, remnants of mergers generate hot gas halos. Here we 
examine the relationship between X--ray luminosity and galaxy age
for a sample of early--type galaxies. 
Comparing normalized X--ray luminosity to three
different age indicators we find that L$_X$/L$_B$ increases with
age, suggesting an increase in X--ray halo mass with time after a galaxy's last
major star-formation episode. The long-term nature of this trend, which
appears to continue across the full age range of our sample, poses a challenge
for many models of hot halo formation. We conclude that models involving
a declining rate of type Ia supernovae, and 
a transition from outflow to inflow of the 
gas originally lost by galactic stars,
offers the most promising explanation for the observed 
evolution in X-ray luminosity.

\end{abstract}

\begin{keywords}
galaxies: interactions -- 
galaxies: elliptical and lenticular -- 
galaxies: evolution --
X-rays: galaxies
\end{keywords}

\section{Introduction}

A potential problem with forming ellipticals from merging spirals 
is how to account for the different gas properties in the two types of 
galaxies. In particular, spirals contain relatively high masses of cold (T
$\sim$ 100~K) gas, whereas ellipticals have very little.
 The opposite situation is true for the hot (T $\sim$ 10$^{6}$~K) gas masses
 -- normal spirals contain rather little hot gas, while ellipticals may 
possess extensive hot halos. Since the total gas masses per unit stellar
mass are broadly comparable in early and late-type galaxies, this raises
the possibility that inefficient merger--induced star formation might
heat the cool interstellar gas in spirals to form the hot halo in
post--merger ellipticals. 
Recently, \scite{gfn2000} have shown that the cold gas mass indeed
decreases in an `evolutionary sequence' from merging spirals to 
post--merger ellipticals. 

However, it is not clear that gas heated in an intense starburst can be
retained within the galactic potential.
\scite{Readponman98} examined the X--ray properties
of eight on--going mergers placed in a chronological 
sequence. 
Their study 
revealed that material is ejected from
merging galaxies soon after the first encounter. Massive extensions of hot
gas are seen (involving up to 10$^{10}$ M$_{\odot}$) at the ultraluminous
peak of the interaction, as the two nuclei coalesce. 
There is evidence for this in in the {\em Antennae}, Arp~220, and 
NGC~2623. However, after this 
phase of peak activity, the X--ray luminosity actually declines. For
example, NGC 7252 (a prime example for the remnant of a merger
that occurred 0.7 Gyr ago)  
shows some evidence of a hot halo, but one that is much smaller
than the halos seen in typical ellipticals.

Two post--merger ellipticals were studied in the X--ray by
\scite{Fabbschweizer95}. They found that neither galaxy had
an extensive hot halo. Given the correlation between X--ray
luminosity and isophotal boxiness (thought to be a signature of a
past merger) demonstrated by \scite{Benderetal89}, this was a
somewhat surprising result. 
If post--merger ellipticals are to resemble `normal' ellipticals
 they need to acquire a hot gas halo. 
Possible mechanisms include:\\
1) The late infall (i.e. after a few Gyr) of HI gas associated
 with the tidal
tails (\pcite{Hibbardetal94}). This cold gas may be shocked to
X--ray
 emitting temperatures as it falls back into the merger remnant.\\
2) A reservoir of hot gas, possibly expelled at 
the nuclear merger stage, might infall from large radii, where its
 low density makes it undetectable to current X--ray satellites.\\
3) After the initial violent starburst, the continued mass loss from 
stars and heating from stellar winds recreates a hot ISM.\\  

The first study to attempt an `evolutionary sequence' for
 post--merger ellipticals was that of \scite{MackieFabb96}.
 For 32 galaxies 
they showed a weak trend for L$_X$/L$_B$ to increase with a
decreasing $\Sigma$ parameter. The $\Sigma$ parameter is defined
by \scite{S&S92}, and is a measure of a galaxy's
fine structure, \ie optical disturbance. They showed that it
correlates with blue colours and Balmer line strength, and is
thus a rough indicator of dynamical youth.
The Mackie \& Fabbiano trend therefore suggested that
post--merger ellipticals became more X--ray luminous, for a given
optical luminosity, as the galaxy aged. 
They also plotted
L$_X$/L$_B$ against
 H$\beta$ absorption line EW, which revealed a similar
trend. A more recent study by \scite{Sansom00} confirms the trend in X--ray 
overluminosity with $\Sigma$ for 38 galaxies, 
and suggests that it might be explained by
the build up of hot gas in post--merger galaxies. However, both
$\Sigma$ and H$\beta$ EW have their drawbacks -- $\Sigma$
 is only semi--quantitative at best and the strength
 of H$\beta$ is affected by both stellar age and metallicity. 

Here we reexamine the trend seen by \scite{MackieFabb96} for a
larger sample, and investigate two new measures of galaxy age: residual
from the Fundamental Plane and spectroscopic age. 
Using these three measures we explore 
the X--ray luminosity evolution of post--merger 
ellipticals. 

Throughout the paper we assume H$_0$ = 75 km s$^{-1}$ Mpc$^{-1}$ and normalise L$_B$ using the
solar luminosity in the B band, L$_{B\odot}$ = 5.2$\times$10$^{32}$ erg s$^{-1}$.

\section{Results}

\begin{figure*}
\centerline{\psfig{figure=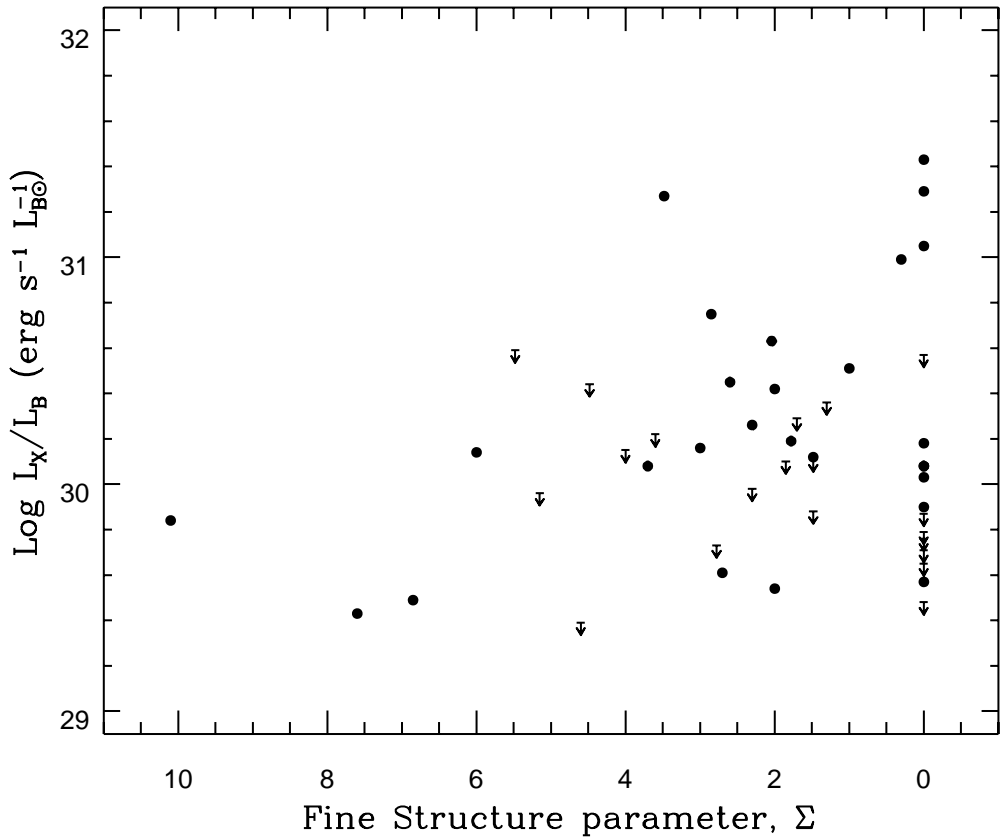,width=6in}}
\vspace{-70mm}
\caption{\label{Sigmafig} Normalized X--ray luminosity versus fine structure
  parameter $\Sigma$ for early--type galaxies. 
Filled circles are detections, arrows denote upper
  limits. Dynamically young galaxies (high $\Sigma$) have low L$_X$/L$_B$.} 
\end{figure*}

We first reexamine the 
trend of normalized X--ray luminosity 
with Fine Structure parameter $\Sigma$,  
presented initially by \scite{MackieFabb96} for 32
galaxies. Here we use a sample of 
47 early--type galaxies with $\Sigma$ taken from \scite{S&S92},
and normalized L$_X$ values from our recent catalogue 
(\pcite{OFP00cat}) 
of X--ray
luminosities (mostly based on {\it ROSAT} data). 
The L$_X$ values are approximately bolometric. 
The L$_B$ values are based wherever possible on B$_T$
magnitudes taken from \scite{pands96}. 
If these are unavailable, values from NED 
are used. Fig.~\ref{Sigmafig} shows the sample of 47 early--type
galaxies plotted in order of 
decreasing $\Sigma$ or increasing age. 

As noted by Mackie \& Fabbiano, the scatter at low values of $\Sigma$ is
large, covering around 2 orders of magnitude in L$_X$/L$_B$, while at
higher $\Sigma$ the scatter appears to be smaller and L$_X$/L$_B$ limited
to low values. A similar trend was seen by \scite{Sansom00} for 38
galaxies. 
This suggests that dynamically young galaxies have low
X--ray luminosities and that aging produces a range of luminosities,
presumably dependent on other factors. 

In order to test the strength of this relation we apply Kendall's K test
to the X--ray detections from our sample. The test does not assume any
distribution in the data, and the K statistic is unit normal distributed
when at least 10 data points are present. For our data we find K =
-2.31054 (using 27 data points), indicating an anti correlation between Log
\lxlb and $\Sigma$ of $\sim$2-$\sigma$ significance. For comparison, we
also apply the test to the sample of Sansom \etal, and find a correlation
of very similar significance, K = -2.28749 for 30 detections.

\subsection{L$_X$/L$_B$ versus Spectroscopic Age}
\label{spec}
\begin{figure*}
\centerline{\psfig{figure=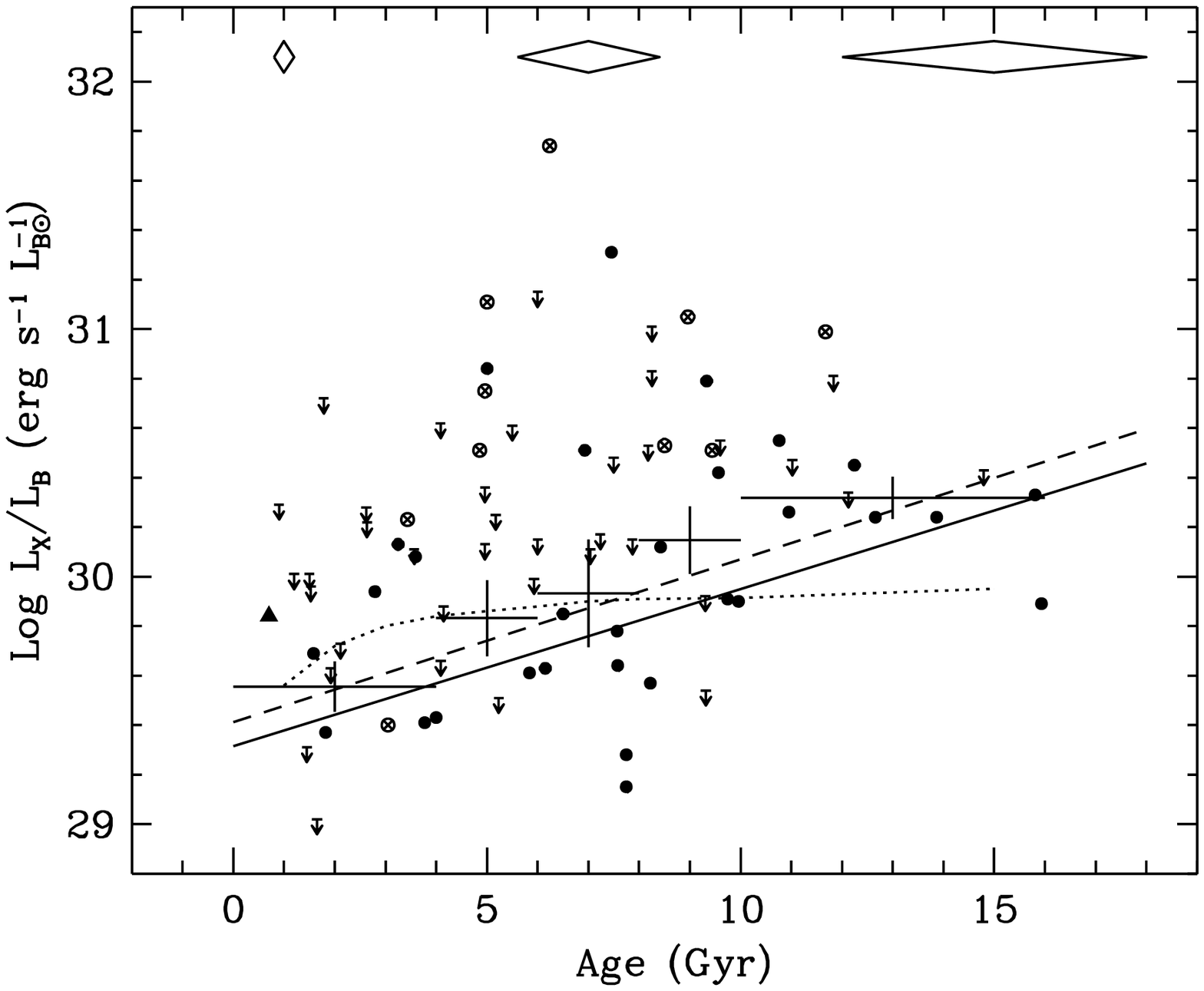,width=6in}}
\vspace{-70mm}
\caption{\label{ageplot} Log (L$_X$/L$_B$) plotted against spectroscopic
  age. Crossed circles represent group or cluster dominant galaxies, filled 
  circles all other detections. Arrows denote upper limits and the filled
  triangle represents NGC 7252, a young post--merger. The three diamonds at 
  the top of the graph show typical age and L$_X$/L$_B$ errors for galaxies
  of age 1, 7 and 15 Gys. The two lines
  are fits to the data, including (dashed line) or excluding (solid line)
  the group and cluster dominant galaxies. The large crosses represent mean 
  L$_X$/L$_B$ for the five age bins described in the text. The dotted line
  shows the expected increase in L$_X$/L$_B$ caused by the decrease of
  L$_B$ with age. Normalisation of this line is arbitrary, and the position 
  shown can be considered to represent a ``worst case'' scenario.}
\end{figure*}

Fine structure only persists in dynamically young galaxies, so
Fig.~\ref{Sigmafig} provides information only on the early galaxy evolution 
of these objects. A more general measure of age is needed to show how
X--ray properties evolve over a longer timescale. Galaxy spectroscopic ages 
are now available for a large number of early--type galaxies from the
catalogue of \scite{TF2000}. These ages are generally based on H$\beta$
absorption line measurements and the stellar population models of
\scite{worthey94}. The line index measurements come from the galaxies'
central regions, and are luminosity weighted. Thus they are dominated by
the last major burst of star formation, which is presumably triggered by
a major merger event. Although not a reliable absolute measure of the age of each
galaxy, these spectroscopic ages do provide us with a much more useful
estimate of their ages relative to one another. 
In Fig.~\ref{ageplot} we show L$_X$/L$_B$ plotted
against spectroscopic age (from \pcite{TF2000}). 

The plot shows a large degree of scatter, most notably in the age range
between 4 and 10 Gyrs. This is likely to be in part caused by the
uncertainties in calculating ages, which we estimate lead a typical error
of $\sim$20\%. We also expect a mean 1-$\sigma$ error in X--ray luminosity
of $\sim$15\%. Typical error bounds for points at 1, 7 and 15 Gys are shown 
in Fig.~\ref{ageplot} by the diamonds at the top of the plot.
The graph also contains a relatively large number of upper
limits, so we have used the survival analysis packages available under IRAF 
to assess the strength of any correlation and fit regression lines. The Cox 
proportional hazard, Spearman's rho and generalized Kendall's tau tests are
used to determine correlation strength. Linear regression fitting is
carried out using the expectation and maximization (EM) algorithm and the
Buckley--James (BJ) algorithm. In all cases we found that the two
fitting algorithms 
agreed closely.

Using the full sample of 77 early-type galaxies, 
we find a correlation of at least
99.6\% significance between L$_X$/L$_B$ and age. Line fits to the sample
produce slopes of 0.066 $\pm$ 0.022 (EM) and 0.063 $\pm$ 0.025 (BJ). The former
is shown as the dashed line in Fig.~\ref{ageplot}. 

Recent work on the X--ray properties of galaxies in groups
(\pcite{Steve}) lead us to suspect that some of the scatter seen
in Fig.~\ref{ageplot} might be caused by inclusion of group or
cluster dominant
galaxies (marked as crossed circles) in our sample. It is also important to 
note that there is evidence to suggest that the largest ellipticals may
have non--solar abundance ratios (\eg \pcite{Carolloetal93}). This could lead to their ages
being underestimated by a small amount, again adding to the
scatter. We therefore removed
from our sample a number of galaxies
listed by \scite{Garcia93} as group
dominant or known cDs and retested the sample. This
improves the correlation slightly (to $>$99.75\%) and makes the line fit
slightly shallower; 0.063 $\pm$ 0.019 (EM) and 0.061 $\pm$ 0.022 (BJ). The EM
fit is plotted as a solid line in Fig.~\ref{ageplot}.

To confirm that the increase in L$_X$/L$_B$ occurs across the range of ages,
rather than just the first few Gyrs, we also binned the sample into subsets 
by age. The age range of each subset was chosen so as to have roughly equal 
numbers of detections in each bin. We then calculated a mean L$_X$/L$_B$ for
each bin, using the Kaplan--Meier estimator. The results are shown in
Table~\ref{bins}, and as large crosses on Fig.~\ref{ageplot}. These
show the same trend for increasing L$_X$/L$_B$ with age, and indicate that
the trend is continuous across the age range covered by the data.

\begin{table*}
\begin{center}
\begin{tabular}{|l|c|c|c|c|}
\hline
Age & No. of detections & No. of upper limits & Log L$_X$/L$_B$ & error \\
(Gyrs) & & & (erg s$^{-1}$ L$_{B\odot}^{-1}$) & \\
\hline
A $<$ 4 & 9 & 12 & 29.56 & $\pm$0.11 \\
4 $\leq$ A $<$ 6 & 6 & 11 & 29.83 & $\pm$0.15 \\
6 $\leq$ A $<$ 8 & 9 & 4 & 29.93 & $\pm$0.22 \\
8 $\leq$ A $<$ 10 & 9 & 6 & 30.15 & $\pm$0.14 \\
A $\geq$ 10 & 8 & 4 & 30.32 & $\pm$0.09 \\
\hline
\end{tabular}
\end{center}
\caption{Mean L$_X$/L$_B$ values for 5 age bins, calculated using the
  Kaplan--Meier estimator.}
\label{bins}
\end{table*}

The Kaplan--Meier estimator (\pcite{Feigelson85}) includes upper
limits by constructing a probability distribution function for the data
in which the probability associated with each upper limit is redistributed
equally over detected values which lie below the upper limit.
So long as there is no systematic difference between the detected and
undetected systems, this should be a reasonable procedure.
A problem arises when the {\it lowest} point in a given bin is an upper limit
(since there are no lower points over which to redistribute the corresponding
probability). When this occurs, the Kaplan--Meier estimator 
treats the limit as a detection and hence may be biased.

Of the five bins chosen, two (the youngest and second oldest) have upper
limits as their lowest values. For the second oldest bin, the problem
data value is
only slightly lower than a detected point, so the bias will be
small. However, the youngest bin has two upper limits as its lowest 
points, one of which is considerably below the nearest detection. To check
for bias, we recalculated a mean log L$_X$/L$_B$ of 29.66 for this 
bin excluding these points, compared to 29.56 previously. 
This indicates that although the youngest bin may be slightly 
biased, it still supports the trend observed.

As a further check of the robustness of our conclusions, we 
examined the  L$_X$/L$_B$:Age relation using detected points only, and
find a correlation significant at $\sim$2.1-$\sigma$. The upper limits also
show a correlation with age (at $\sim$2.3-$\sigma$). These trends
cannot be selection effects in the data, since neither source
distance, nor X-ray exposure time are correlated with age. They
therefore point to a genuine trend in the data. Galaxies with larger
spectroscopic ages typically have higher L$_X$/L$_B$ than younger ones.

As the merger--induced starburst fades, we would expect the blue 
luminosity to decline with time. This will cause an increase in 
L$_X$/L$_B$, even for constant L$_X$, which will combine with any trends in 
L$_X$ which are present. With this is mind, we show in Fig.~\ref{ageplot} the  
expected change due to a fading starburst from stellar population models 
(Worthey 1994). Here we have crudely assumed that the progenitor galaxies 
consisted of a solar metallicity, 15 Gyr old population, and that the new 
stars created in the merger also have solar metallicity. The relative 
mass ratios are 90\% progenitor stars and 10\% new stars. The total B 
band luminosity fading of this composite stellar population is shown in 
Fig. 2 (we assumed that the fading of the progenitor stars after reaching 18 
Gyrs old is insignificant). It can be seen 
that the starburst fades rapidly in the first few Gyrs, roughly matching 
the trend seen in L$_X$/L$_B$, but at later times there is very little 
change, whereas the observed mean L$_X$/L$_B$ value continues to rise. So
although the L$_X$/L$_B$ evolution at early times could be simply driven by
the fading starburst, this cannot account for the evolution at late times. 

\begin{figure*}
\centerline{\psfig{figure=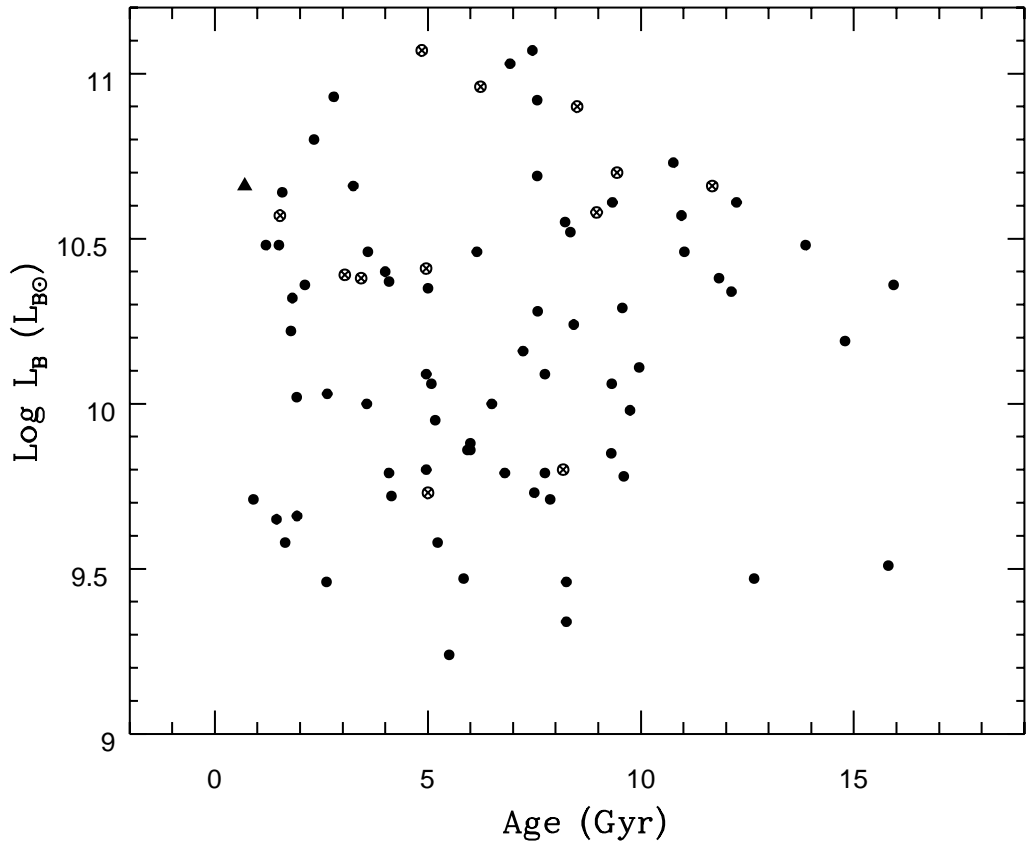,width=6in}}
\vspace{-70mm}
\caption{\label{Lb} Optical luminosity versus spectroscopic
  age. Crossed circles represent cluster or group dominant galaxies
  and filled circles all other detections. The triangle is the recent merger
remnant NGC 
  7252. The cE galaxy, M32, 
has been excluded from the plot and associated statistical tests.}
\end{figure*}

A further possibility to be checked is that our sample might contain a
trend in mean optical luminosity with age. Since L$_X$/L$_B$ has been
widely reported (see however \scite{Steve}) to rise with L$_B$, this
could lead to a consequent correlation between age and L$_X$/L$_B$.
To test this, we show in Fig.~\ref{Lb} a plot of log L$_B$ with measured
age. An anti--correlation between L$_B$ with age could produce the observed
trend in L$_X$/L$_B$. Fig.~\ref{Lb}
does not appear to show any such trend, and in order to
test for a correlation, we apply Kendall's K test to the data. We find a K
statistic (which is unit normal distributed) of $\sim$0.44 (for 77
galaxies), and excluding group dominant galaxies reduces this to
$\sim$0.33 (for 65 galaxies). This result indicates that there is no trend in 
the data, even at the 1$\sigma$ level. 
We therefore conclude that our sample does not show an
anti--correlation of L$_B$ with age, and that the L$_X$/L$_B$ trend
with age reflects an increasing X--ray luminosity.

\subsection{L$_X$/L$_B$ versus Fundamental Plane Residual}
\begin{figure*}
\centerline{\psfig{figure=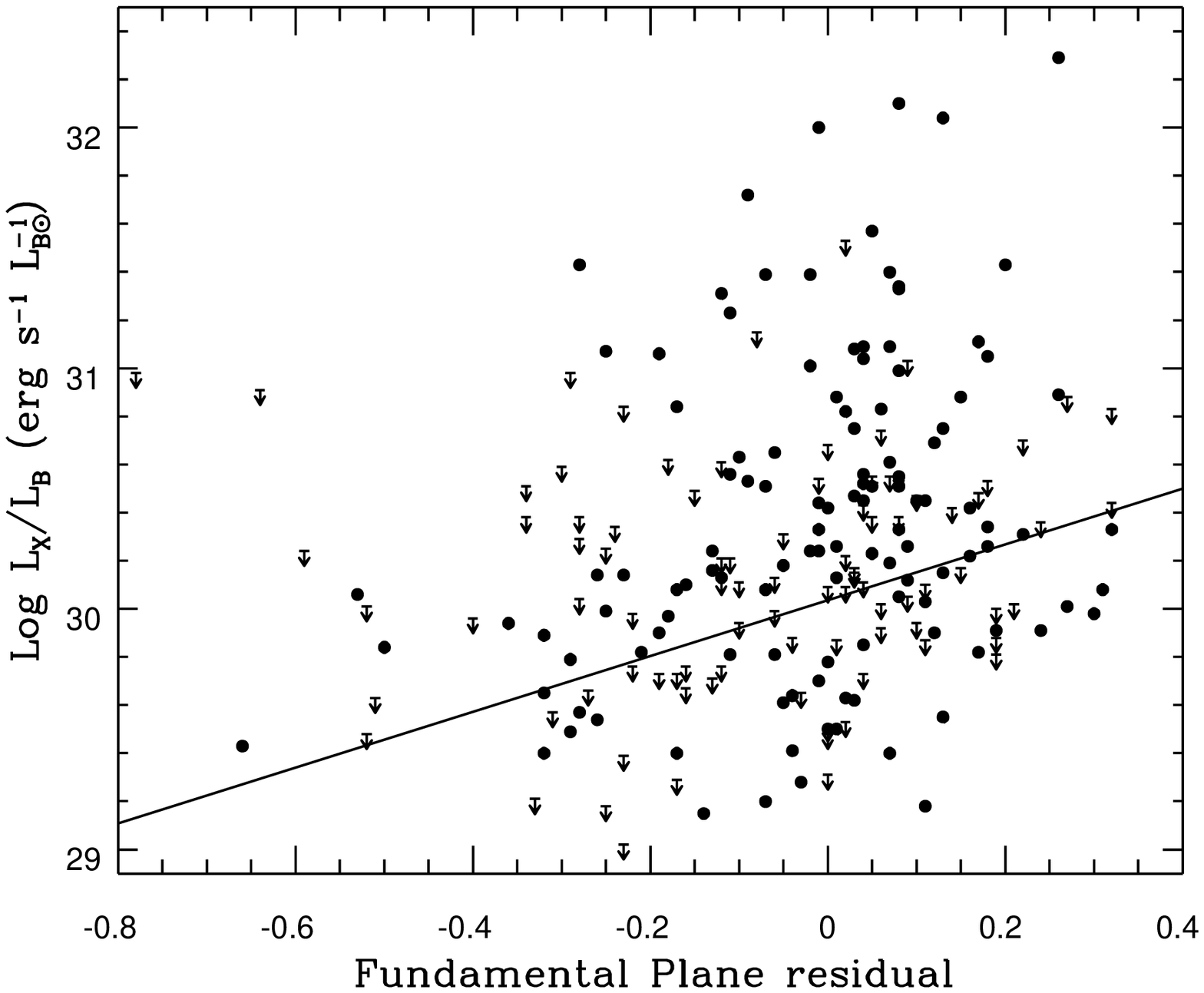,width=6in}}
\vspace{-70mm}
\caption{\label{FPres} X--ray luminosity versus Fundamental
  Plane residual for early--type galaxies. Filled circles represent detections and arrows upper
  limits. The solid line is a fit 
  to all points with --0.5$<$FP residual$<$0.5}
\end{figure*}

As further evidence of X--ray evolution with age we have compared
L$_X$/L$_B$ to residual from the Fundamental Plane (FP). Work by
\scite{forbesponmanbrown98} has shown that galaxy age appears to be a ``fourth 
parameter'' affecting the position of a galaxy relative to the FP
plane. Galaxies below the plane (negative residual) are generally young,
while older objects lie on or above the plane (positive 
residual). Fig.~\ref{FPres} shows 212 galaxies with Fundamental Plane
residuals taken from \scite{pands96}.

As with Figs.~\ref{Sigmafig} and ~\ref{ageplot}, this shows a trend of
increasing L$_X$/L$_B$ with age. For the complete data set, the correlation 
strength is $>$99.98\%. However, only a small number of points lie outside the
range --0.5$<$FP residual$<$0.5. These are unlikely to be representative of
the general population, but will have a strong influence on the statistical 
tests. Excluding them lowers the correlation strength to
$\sim$99.95\%, with a slope of 1.39 $\pm$ 0.40 (EM). This line is shown 
in Fig.~\ref{FPres}. Again the trend shown in Fig.~\ref{FPres} is
one of increasing X--ray luminosity, relative to the optical, as
a galaxy evolves.

\section{Discussion}
In the previous section we have used three age estimators to show a strong
correlation between normalized X--ray luminosity and galaxy age. The X--ray 
emission from early--type galaxies is known to be produced by sources of
two main types; discrete sources such as X--ray binaries, and hot
gas. The contribution of discrete sources has been shown to be important in
low luminosity early-type galaxies (\pcite{fkt94};
\pcite{irwinsarazin98a}; \pcite{SarazinIrBreg00}) and in late-type
galaxies, where they are generally the dominant source of X--ray emission. On the
other hand, most early--type galaxies have a strong hot gas component, and
massive ellipticals are certainly dominated by hot gas emission
(\pcite{Matsushita00}; \pcite{Matsushita00b}). As the contribution of hot gas 
is known to vary a great deal, while the contribution from discrete sources 
is generally believed to scale with L$_B$ (\pcite{Matsushita00};
\pcite{fabbianokimtrinchieri92}) , it seems likely that 
the trend we observe in L$_X$/L$_B$ is caused by changes in the hot 
gas content of these
objects with age. We now go on to discuss possible hot halo formation
mechanisms which may be responsible for this trend of increasing 
X--ray gas content with age. 

\subsection{Gas infall}
It has been suggested (\pcite{HibvanG96}) that the hot gas in elliptical
galaxies may be 
produced during a merger by the shock heating or photoionization of cool
gas from the progenitor galaxies. In some ongoing mergers the tidal tails
are thought to contain up to half the \hi gas originally present in 
the progenitor galaxies (\pcite{Hibbardetal94}). When this gas falls
back into the body of the galaxy, shock heating should be capable of
heating it to X--ray temperatures. An alternative is that the
temperature of the gas is caused by heating in the starburst phase of 
the merger. In this case the hot gas would be blown out to large radii, but 
might be contained by the dark matter halo of the galaxy
(\pcite{MathBrig97}). At these large radii it would be too diffuse to be
detected, but would eventually fall back into the galaxy, forming the
observed X--ray halo. 

In terms of available masses of gas, both these models appear to be viable
formation processes. \scite{Bregman92} derived X--ray and cold gas masses
for a large 
sample of early--type galaxies, finding M$_X$ of between $\sim 10^6$ and
10$^{11}$ M$_\odot$. The mean X--ray gas mass was a few 10$^9$
M$_\odot$. On the other 
hand, very few detections of \hi were made in early--type
galaxies, although Sa
galaxies  
were found to contain between 10$^7$ and $\sim 5 \times 10^{10}$ M$_\odot$
of \hi. Studies  
of \hi masses in later--type galaxies (\pcite{Huchtmeier89};
\pcite{Huchtmeier88}) give an average of a few 
10$^9$ M$_\odot$, depending on the range of morphologies chosen.
Given that large elliptical galaxies are likely to have been formed by the
merger of many smaller spiral galaxies, these results suggest that fairly
modest conversion efficiencies could produce the expected X--ray halo gas masses
either from infalling \hi or via starburst heating and later infall.

However, the infalling \hi model fails to explain the timescale over
which we observe the L$_X$/L$_B$ trend. In the well known merging galaxy NGC
7252, 50\% of the cold gas 
currently in the tidal tails is expected to fall back into the main body of 
the galaxy within the next 2--3 Gyr (\pcite{HibvanG96}). This suggests that
we should expect to see a rapid build up of X--ray emission in the first
few gigayears after merger. This increase should be made even more
noticeable by the fading of the stellar population, as this is the period
during which L$_B$ is dropping most quickly. Rapid generation of the X--ray
halo is inconsistent with our results.

A model involving infall of hot gas is likely to have similar problems. It is
difficult to see how gas falling back into the body of the galaxy can do so 
at the steady rate needed to produce a smooth increase in
L$_X$/L$_B$. Infall may be delayed, as the hot gas reaches larger radii
than the cold, but it will still occur on a galaxy dynamical
timescale. Models involving cooling and infall of gas onto central dominant 
galaxies in groups or clusters (\pcite{BrigMath99}) may be able to produce
an inflow over a longer period, but these are unlikely to be generally
applicable. Group dominant galaxies have exceptionally large 
dark halos, making them more able to retain (or accrete) hot gas. They also 
lie at the bottom of a group potential well, and may be at the centre of a
group X--ray halo, providing them with an extra reservoir of hot gas. These 
conditions do not apply to the majority of early-type galaxies. Therefore
we suggest that infall of (cold or hot) gas is unlikely to be the dominant
process in the generation of X--ray halos in typical ellipticals. 

\subsection{Ongoing stellar mass--loss and galaxy winds}
Since infall of either hot or cold gas seems unable to provide the sort
of long term trend in L$_X$/L$_B$ seen in Fig.~\ref{ageplot}, we now look to
models involving gas generated by mass loss from the galactic
stars.
This source of gas is fairly well understood, and can provide the
sort of gas masses required to account for the halos of many early-type
galaxies (though probably not the very brightest -- \pcite{Brighenti98}), 
provided that the gas is retained 
within the galactic potential. Two main timescales are
involved. One is the mass
loss rate from stars, primarily giant stars. The second is the rate of type Ia 
supernovae (SNIa) which provide the main 
heat source (after the brief SNII phase).
These two factors, and the interplay
between them, can potentially lead to changes in the hot gas content of
elliptical galaxies on timescales much longer than a dynamical time,
and so have the potential to explain the slow trend we observe.
The stellar mass loss rate is fairly well understood (e.g. \pcite{Ciotti91}),
and for a single-aged stellar population it declines approximately as 
$t^{-1.3}$. However, the SNIa heating rate, which dominates the effective
specific energy of the injected gas, is highly uncertain. Recent estimates
of the SNIa rate in old stellar populations (e.g. \pcite{Cappellaroetal99}) have 
revised the classic \scite{Tammann82} value of the rate in old
stellar populations down by a factor 3-4, and the evolution of this
rate with population age is very model-dependent. Given our continuing
ignorance about the precise nature of SNIa, it must therefore be regarded as
largely unknown.

The specific energy of the injected gas determines whether gas is retained
in a hydrostatic hot halo, or escapes the galaxy as a wind. The way in which
the specific energy evolves, depends upon the evolution of the SNIa rate
relative to the stellar wind loss rate. There are therefore two fundamentally
different classes of models: those in which the 
SNIa rate is constant, or declines
more slowly than the stellar mass loss rate (e.g. 
\pcite{Loewenstein87,David91})
and those in which the supernova rate drops more quickly than the gas
injection rate (\pcite{Ciotti91,Pellegriniciotti98}). Broadly speaking, when
the specific energy of the injected gas is smaller than its gravitational
binding energy it will be retained in a hot halo. It may
subsequently cool in a luminous cooling flow. However, if the specific
energy of the gas substantially exceeds its binding energy it will
escape from the galaxy in a fast wind, which results in a much
lower X--ray luminosity. 
The first class of models, in which the specific energy rises with time,
therefore produces an evolution from an early inflow phase to a later wind
phase, predicting a {\it declining} X-ray luminosity with time. Such models are
clearly inconsistent with our results.

Models in which the SNIa rate drops more quickly than $t^{-1.3}$ evolve, in 
broad terms, from a low luminosity wind phase, towards a more luminous
halo/cooling flow phase. The timescale for this evolution depends upon
a variety of factors, such as the depth and shape of the potential,
but it is typically $\sim 10$~Gyr (\pcite{Ciotti91}). In fact, none of
the models studied to date shows the rather simple monotonic rise
in L$_X$ with age which we observe. The models of \scite{Ciotti91} predict
an initial luminous phase, when the gas loss rate from stars is very high.
The X--ray luminosity 
then declines as the wind loss rate drops, and then rises again
during the transition to a bound hydrostatic halo, at which point
the X-ray luminosity is essentially equal to the SNIa luminosity. 
This whole development describes a fairly symmetrical dip and rise, 
lasting $\sim5-15$~Gyr, which is again 
not what we see in the data. However, this model describes
a galaxy in which {\it all} stars are formed at $t=0$. In contrast, 
the spectroscopic age of our
galaxies probably denotes the time since a merger-induced starburst
involving only a few percent of the stellar mass. In this case, the early
stellar mass loss rate will be much lower than in the Ciotti et al models,
and the X-ray luminosity correspondingly less (scaling approximately
as the square of the mass loss rate). The main change in L$_X$ will be a
rise associated with the slowing wind and transformation 
into a hydrostatic halo. In addition, as shown in Fig.~\ref{ageplot},
the decrease in optical luminosity 
as the starburst population fades will produce a 
rise in L$_X$/L$_B$ over the first 1-2~Gyr.

\section{Concluding Remarks}

We have examined the evolution of the X--ray properties of
early--type galaxies. For 
three galaxy age estimators (fine structure parameter, Fundamental
Plane residuals and spectroscopic ages) we find that the
normalized X--ray luminosity evolves with time. 
In particular, the X--ray luminosity, which reflects the mass of
hot halo gas, appears to increase at a steady rate over $\sim$ 10
Gyrs.

Comparing the long term trend in L$_X$/L$_B$ which we observe with
expectations from 
possible mechanisms for hot halo formation, we conclude that the
only viable mechanism appears to be the slow evolution from an
outflowing 
wind to hydrostatic halo phase driven by a declining SNIa rate. 
Infalling gas seems unlikely to be the main cause of such a long
term trend. Our results suggest that some of the scatter seen in
the global L$_X$ versus L$_B$ relation is due to the evolutionary
state, and past merger history, of early--type galaxies.

\vspace{0.5cm}
\noindent{\bf Acknowledgements}\\
We would like to thank R. Brown for early work on this project and
S. Helsdon, W. Mathews and A. Renzini for useful discussions. We also thank 
an anonymous referee for several suggestions which have improved the
paper. This research has made use 
of the NASA/IPAC Extragalactic Database (NED) which is operated by the
Jet Propulsion Laboratory, California Institute of Technology, under
contract with the National Aeronautics and Space
Administration. E. O'S. acknowledges the receipt of a PPARC studentship.\\

\bibliographystyle{mnras}
\bibliography{../paper}

\label{lastpage}
\end{document}